# Assessing the Performance of Recent Density Functionals for Bulk Solids

Gábor I. Csonka,[1] John P. Perdew,[2] Adrienn Ruzsinszky,[2] Pier H. T. Philipsen,[3] Sebastien Lebègue,[4] Joachim Paier,[5] Oleg A. Vydrov,[6] and János G. Ángyán[4]

[1] *Department of Inorganic and Analytical Chemistry, Budapest University of Technology and Economics, H-1521 Budapest, Hungary*
[2] *Department of Physics and Quantum Theory Group, Tulane University, New Orleans, Louisiana 70118*
[3] *Scientific Computing & Modelling NV, Vrije Universiteit, Theoretical Chemistry, De Boelelaan 1083; 1081 HV Amsterdam; The Netherlands*
[4] *CRM2, UMR 7036, Institut Jean Barriol, Nancy-University and CNRS, B. P. 239, F-54506 Vandoeuvre-lès-Nancy, France*
[5] *Faculty of Physics, Universität Wien and Center for Computational Materials Science, Sensengasse 8/12, A-1090, Wien, Austria*
[6] *Department of Chemistry, Massachusetts Institute of Technology, Cambridge, Massachusetts, 02139*

We assess the performance of recent density functionals for the exchange-correlation energy of a nonmolecular solid, by applying accurate calculations with the GAUSSIAN, BAND, and VASP codes to a test set of 24 solid metals and non-metals. The functionals tested are the modified Perdew-Burke-Ernzerhof generalized gradient approximation (PBEsol GGA), the second-order GGA (SOGGA), and the Armiento-Mattsson 2005 (AM05) GGA. For completeness, we also test more-standard functionals: the local density approximation, the original PBE GGA, and the Tao-Perdew-Staroverov-Scuseria (TPSS) meta-GGA. We find that the recent density functionals for solids reach a high accuracy for bulk properties (lattice constant and bulk modulus). For the cohesive energy, PBE is better than PBEsol overall, as expected, but PBEsol is actually better for the alkali metals and alkali halides. For fair comparison of calculated and experimental results, we consider the zero-point phonon and finite-temperature effects ignored by many workers. We show how Gaussian basis sets and inaccurate experimental reference data may affect the rating of the quality of the functionals. The results show that PBEsol and AM05 perform somewhat differently from each other for alkali metal, alkaline earth metal and alkali halide crystals (where the maximum value of the reduced density gradient is about 2), but perform very similarly for most of the other solids (where it is often about 1). Our explanation for this is consistent with the importance of exchange-correlation nonlocality in regions of core-valence overlap.

# I. INTRODUCTION

Popular or standard generalized gradient expansion approximations (GGA)[1] for the exchange-correlation energy of Kohn-Sham density functional theory[2,3] improve upon the local density approximation (LDA) for atomization energies of molecules and enthalpies of formation derived from atomization energies, but GGAs (e.g., PBE[1]) do not improve the calculated lattice constants for typical nonmolecular solids. Phonon frequencies, ferromagnetism, ferroelectricity, and many other properties are critically volume-dependent, and thus highly accurate lattice parameters are indispensable for these properties. Neither LDA nor GGA is clearly to be preferred for solid state applications, both giving errors of comparable magnitude (though generally of opposite sign). At all temperatures, LDA systematically underestimates lattice constants and coefficients of thermal expansion, whereas GGA overestimates them.[4] In contrast, LDA (GGA) overestimates (underestimates) bulk moduli and phonon frequencies. This behavior is a well established trend observed in numerous previous studies.[5] For a small set of 9 metals, Grabowski et al.[5] have found that LDA underestimates the lattice constant on average by −0.7% and GGA overestimates it on average by 1.8%. Generally, the error in the bulk modulus, is much larger in magnitude (LDA average, 11.6%; GGA average, −13.7%) and inversely related to the error in the lattice constant. The inverse relation can be explained by the volume dependence of the total energy causing a monotonous decrease of the equilibrium bulk modulus $B_0$ with increasing equilibrium volume.[5] Recently a linear combination of GGA and LDA results was applied (with 0.57 and 0.43 coefficients, respectively) for the equilibrium lattice constant of Al.[6]

Popular GGAs (e.g., PBE[1] and B88[7] + GGA correlation) fail seriously for the exchange–correlation component of the jellium surface energy, while LDA performs surprisingly well in that case. A detailed analysis of the exchange-correlation components shows that LDA benefits from large error compensation. It has been observed[8,9] that in GGAs this delicate balance between exchange and correlation is not valid any more, although exchange and correlation components of the surface energy are separately improved.[8,9] A recent study[10] shows that even if the PBE constraints are maintained, they can be satisfied by a continuous range of diminished gradient dependence (DGD) GGAs lying between PBE and LDA. In DGD GGAs, a balanced error cancellation between exchange and correlation is restored, which in turn results in good surface energies.

Meta-GGAs using the positive kinetic energy density (like TPSS[11], which adds several more exact constraints to those satisfied by PBE) might give excellent jellium surface energies, but do not improve sufficiently upon the lattice constants predicted by standard semilocal approximations,[9] although TPSS improves molecular atomization energies[12] and many other properties.[13]

Following the realization that popular GGA and meta-GGA functionals fail for solid properties (e.g., for the energy-volume equation of state, surface energies, etc.), recently many modified functionals (all GGAs) have appeared in the literature providing improved results for solids at the expense of worsening the atomic total and molecular atomization energies.[14,15,16] Kohn and Mattsson[17] proposed an alternative approach for incorporating effects of the inhomogeneity of the electron density: the Airy gas approximation, a description of the electronic edge within the linear potential. Vitos et al.[18] constructed from that model a GGA for exchange (the local Airy gas or LAG functional) by fitting to the Airy gas conventional exchange energy density, and this exchange was combined with LDA correlation. In the absence of an Airy gas correlation energy density, the AM05 condition,[19] fitting a density functional to the jellium surface exchange-correlation (xc) energy, was used to construct the AM05 GGA[19] for correlation.

The exchange gradient expansion coefficient ($\mu$) of the popular GGAs was set to obtain good atomic and atomization energies and good thermochemistry (where enthalpies of formation are traditionally calculated via atomization energies, not from the calculated energies of the standard states of the elements). This coefficient, however, is about twice as large as for the exact slowly-varying gradient expansion coefficient for exchange ($\mu = 0.22\text{-}25$ vs. $10/81 \cong 0.12346$).[14] The PBE GGA can be rebiased toward solids and surfaces by changing the exchange as well as the correlation gradient coefficient.[14] Recovering the slowly-varying gradient expansion for exchange for a wider range of the reduced density gradient $s$ (as defined in section II), combined with the jellium surface energy condition[19] for correlation[14], leads to a revised PBE GGA for solids, named PBEsol GGA.[14] Nonmolecular solids have important valence regions over which the density variation is so slow (with reduced gradient $s < 1$) that the exchange energy can be described by the second-order gradient expansion. This suggests that recovery of the second-order gradient expansion over this range of $s$ is a relevant constraint on a generalized gradient approximation for exchange in solids (although a similar constraint is not so relevant for correlation).[20] Importantly, the new PBEsol performs accurately for the exchange component of the jellium surface energy, not relying on a fit to the latter. Furthermore, PBEsol outperforms the original PBE GGA by correctly predicting the energy differences between isomers of hydrocarbons,[21] while most of the GGAs and meta-GGAs fail for this long-standing problem.[21] The larger gradient coefficient for exchange in the original PBE GGA is needed to produce the correct asymptotic expansion of the exchange energy for a neutral atom of large atomic number Z, as shown in Ref. 20. In this large-Z limit, the electron density becomes slowly-varying over space, except near the nucleus and in valence and tail regions.[20] Under an appropriate scaling, atomic densities can become slowly-varying essentially everywhere.[20]

The performance of PBEsol was studied in several recent papers. Ropo et al.[22] compared the performance of the PBEsol,[14] PBE,[1] AM05,[19] and LAG[18] approximations for 29 metallic bulk systems (mono- and divalent *sp*, plus several 3*d*, 4*d*, and 5*d* metals). These calculations were performed using the exact muffin-tin orbitals (EMTO) method. The EMTO method is a screened Korringa-Kohn-Rostoker method that uses optimized overlapping muffin-tin potential spheres to represent the one-electron potential. The applied method has a limited precision (about 0.01 Å for the lattice constant, and 4 GPa for the bulk modulus), and the calculations were compared to uncorrected experimental results (i.e., the lattice constants and bulk moduli measured at 300 K were used for many metals, and zero-point phonon effects were ignored for all metals). Nevertheless, the aforementioned study confirms the good performance of the PBEsol,[14] LAG[18] and AM05[19] functionals for most of the metals, except for the 3*d* metals. For these metals PBE agrees better with the uncorrected experimental results. For most 3*d* metals even the PBE functional gives too-small lattice constants, and thus the even shorter lattice constants given by PBEsol worsen the agreement. (Note that LDA results are quite poor for these metals).

Other recent studies of PBEsol have also been made. The PBEsol functional predicts correctly the 2D-3D shape transition for gold clusters.[23] It was tested recently for the compression curves of 8 transition metals (Fe, Co, Ni, Zn, Mo, Ag, Pt, and Au) in the Mbar pressure range.[24] It was found that PBEsol gives an equation of state (EOS) closer to experiment than PBE for Mo, Ag, Pt, and Au, although the overall accuracy of the PBE is somewhat better (due to the more accurate ε-Fe results).[24] PBEsol is expected[14] to become more accurate as a solid becomes more compressed under pressure. We believe we can see evidence for this in Fig. 4 of Ref. 24, even for the 3$d$ transition metals. PBEsol was applied to the B1 rock-salt-type phase of metallic thorium carbide,[25] and with considerable success to the structural, electronic, and phonon properties of the cubic and tetragonal phases of SrTiO$_3$ and BaTiO$_3$.[26] PBEsol was also tested in a general discussion of material simulations.[27] The PBEsol of course shares the limitations of all GGA functionals.[14,28,29,30] Several comments and replies on the GGAs for solids have appeared recently.[31,32,33,34] Contrary to what might be inferred from Refs. 33 and 34, no GGA can recover the correct fourth-order gradient expansion for the exchange energy, even approximately, but a meta-GGA can (and TPSS in fact does, at least for very slowly-varying densities).

The validation of Kohn-Sham xc-functionals[2,3] can become particularly dubious if relatively low-precision theoretical calculations are compared with experimental data with sizable uncertainties, e.g., due to the lack of thermal and anharmonic expansion corrections in our case. The present work compiles highly accurate anharmonic-expansion-corrected experimental results and compares them with results obtained using methods based on either Gaussian-type orbital (GTO) basis sets as implemented in GAUSSIAN,[35] numerical atomic orbital (NAO) and Slater-type orbital (STO) basis sets as implemented in BAND[36] (BAND/linear combination of atomic orbitals, LCAO), or projector augmented plane waves (PAW) as implemented in VASP (VASP/PAW).[37] Moreover, we present a suitable methodology for testing density functionals for solids and revisit previous results to be found in the literature.

We use a test set of metals (main-group and transition metals) and non-metals (semiconductors and ionic insulators) comprising 18 solids compiled by Staroverov et al.[38] The test set contains four main-group metals (Li, Na, K, Al), four transition metals (Cu, Rh, Pd, and Ag), five covalent solids (diamond, Si, $\beta$-SiC, Ge, GaAs), and five ionic solids (NaCl, NaF, LiCl, LiF and MgO). This test set was extended by six more main group metals (Rb, Cs, Ca, Sr, Ba, Pb). All solids were calculated in their ambient-condition crystal structures and non-magnetic phases.

## II. DEFINITIONS

Here we summarize some standard definitions used in this article. Consider a solid in which the total energy per atom is $e$, and the volume per atom is $v$. We can compute a binding energy curve $e(v)$. The equilibrium volume $v_0$ minimizes $e(v)$:

$$\frac{de}{dv} = 0 \qquad (v=v_0). \qquad (1)$$

The bulk modulus is related to the second derivative at the minimum:

$$B_0 = v\frac{d^2e}{dv^2} \qquad (v=v_0). \qquad (2)$$

The cohesive energy is the energy per atom needed to atomize the crystal:

$$e(\infty) - e(v_0). \qquad (3)$$

Measurements of these quantities include the effects of nuclear vibration, while density functional calculations give most directly the values for a static lattice.

A GGA for the exchange-correlation energy can be written as

$$E_{xc}[n_\uparrow, n_\downarrow] = \int d^3r \; n \cdot \varepsilon_{xc}^{unif}(n) \cdot F_{xc}(s, r_s, \zeta). \qquad (4)$$

Here $n_\uparrow$ and $n_\downarrow$ are the electron spin densities, $n = n_\uparrow + n_\downarrow$ is the total density, and $\varepsilon_{xc}^{unif}(n)$ is the exchange-correlation energy per particle of a spin-unpolarized electron gas of uniform density $n$. In atomic units (hartrees),

$$\varepsilon_{xc}^{unif}(n) = \frac{-3}{4\pi \cdot r_s}\left(\frac{9\pi}{4}\right)^{\frac{1}{3}} \qquad (n = 3/[4\pi \cdot r_s^3]). \qquad (5)$$

The enhancement factor $F_{xc}$, which distinguishes one GGA from another, depends also upon the relative spin polarization $\zeta = (n_\uparrow - n_\downarrow)/n$ (which vanishes in our solids at equilibrium, but not typically in their free atoms) and on the reduced density gradient

$$s = \frac{|\nabla n|}{2\cdot(3\pi^2 n)^{1/3} n}, \qquad (6)$$

which expresses how fast the density varies on the scale of the local Fermi wavelength $\lambda_F = 2\pi/(3\pi^2 n)^{1/3} = 3.274\, r_s$. The exchange enhancement factor $F_x$ does not depend upon $r_s$ (and in fact is the $r_s \to 0$ limit of $F_{xc}$). Plots of the enhancement factors provide a way to visualize the $s$−dependence of the GGA. When $s$ is set to zero, a GGA reduces to LDA.

### III. METHODS

In our previous studies[14,38] we used GTO basis sets developed for atomic and molecular calculations.[39] This kind of basis set frequently includes small-exponent (less than 0.10) diffuse functions that are far reaching. Inclusion of diffuse functions into a GTO basis set frequently improves the DFT results for molecules.[40] However, diffuse functions decay very slowly with distance and slow down dramatically the calculation of Coulomb contributions to the total energy of crystals. For crystals the standard GTO basis sets have to be modified as described in our earlier papers.[14,38] The GTO basis-set incompleteness limits the accuracy of the calculated lattice constant to 0.03 Å for metals; however, for covalent, semiconductor and ionic solids, carefully modified Gaussian basis sets might perform quite well.[41] We compare our results to basis sets denoted by GTO1 used in Ref. 38 and GTO2 used in Refs. 14 and 41. The two basis sets are different for C (diamond), Si, SiC, Ge, GaAs, and MgO.

As demonstrated in Ref. 42, PBE equilibrium lattice constants obtained using PAW (VASP, Ref. 37) and full-potential linearized augmented plane wave (FP-LAPW, WIEN2k[43]) methods are *de facto* identical. In addition, comparing those PBE lattice constants to the ones obtained using the LCAO code BAND, one realizes that BAND results compare very well to the results obtained using the aforementioned codes (see Table I). Hence, those codes give consistent results, which are free from the problems arising when Gaussian basis sets are used for extended systems. In this work, we compare our earlier results[14,38] calculated using a modified version of the GAUSSIAN program[35] with new results calculated using BAND and VASP. Our PBE results from VASP can be directly compared to those of Paier *et al.*[42] The discrepancies are small and caused by slight differences in the volume range governing the Murnaghan fits. Importantly, none of the differences affect any conclusions.

The VASP calculations presented in this work are based on the PAW,[44,45] which describes the electron-ion interaction. Characteristics of PAW are i) the inclusion of effects of the nodal structure of valence wave functions close to the ionic cores and ii) the preservation of the orthogonality between the valence and the core states. Note that the chemically inert core states are usually kept frozen, but this is not inherent to PAW. For a profound description of an all-electron (i.e., no frozen cores) implementation of PAW into VASP, we refer the reader to the literature.[46] Note that all PAW core potentials include scalar relativistic corrections. At this point, the authors wish to briefly discuss two issues: First, the precision of the frozen-core PAW implementation of VASP has been thoroughly tested against the all-electron full-potential linearized augmented plane wave (FP-LAPW) plus local orbitals (lo) method (WIEN2k[43]), which is commonly regarded as the benchmark method for solid state applications. For a test set comprising main group metals (Li, Na, Al), *d*-metals (Cu, Rh, Pd, Ag), as well as semiconducting and ionic insulators (C, GaAs, MgO), the agreement between PAW and FP-LAPW+lo results (lattice constants, bulk moduli) is excellent (see Sec. B.1 of Ref. 42). Second, it is possible to use multiple xc-functionals on the same set of PAW core potentials without sacrificing the high precision.[47] Possible transferability errors are largely reduced, if not eliminated, by virtue of the consistent recalculation of the core-valence interaction with the selected density functional. Although the core states are frozen in the configuration determined as the PAW core potential is generated using a density functional which might differ from the selected one, the errors thereby introduced are insignificant (e.g., LDA PAW core potentials, combined with the PBEsol xc functional in an actual application; for more details see Sec. III of Ref. 48). The PAW pseudopotentials we have used are summarized in the Supplementary Material. The technical specifications to the VASP calculations read as follows: for the PBEsol calculations, a kinetic energy cut-off of 500 eV was used, except for Li (600 eV). All Brillouin zone integrations were performed on Γ-centered symmetry-reduced Monkhorst-Pack[49] *k*-point meshes, using the tetrahedron method with Blöchl corrections.[50] For Li (20x20x20) *k* points and for the remaining solids (16x16x16) *k* points were used. As outlined in Sec. III of Ref. 42, this setup ensures that the results are converged to within all specified digits. In the calculations for K and Ge presented in Tables V and VI of this work, (24x24x24) *k* points were used. A plane-wave cutoff of 600 eV was applied to K and a cutoff of 750 eV was applied to Ge. To minimize errors arising from the frozen core approximation, we used PAW data sets treating the K 3*s* and 3*p* states and the Ge 3*d* states as valence electrons.

BAND LCAO calculations were performed at a benchmark level with the finest grid available, together with a very dense *k*-space sampling (keywords in BAND: accuracy 6, Kspace 7), using the LDA electron densities. In other words, the BAND results for the beyond-LDA functionals are not fully selfconsistent and demonstrate that full selfconsistency is not needed for high accuracy. We use the large QZ4P basis set consisting of numerical atomic orbital (NAO) core orbitals and one NAO plus three Slater-type orbitals (STOs) for the valence functions. The core is kept frozen during the self-consistency loop and very small in order to eliminate any significant effects of this approximation. For a discussion of errors in BAND/LCAO calculations, we refer the reader to Ref. 51. We estimated the effect of self-consistency using the TZ2P basis set and found a 0.002 Å effect on the lattice constant and a 0.5% effect on the bulk moduli. The relativistic calculations were performed within the zeroth-order regular approximation (ZORA)[52], an accurate approximation to the Dirac equation. For details of the implementation, see Ref. 53. We checked the spin-orbit effects, and found them negligible for the solids in this study.

We estimated the equilibrium lattice constant $a_0$, bulk modulus $B_0$, and pressure derivative $B_1$ at T = 0 K by calculating the energy of the unit cell at 7-15 points in the range $v_0 \pm 5\%$ (where $v_0$ is the equilibrium volume per atom), then fitting the data to analytic equations of state (EOS) $e(v)$. The relation between the lattice constants and the mono-atomic cell volumes is: $v_0 = a_0^3/2$ for the A2(bcc) crystal, $v_0 = a_0^3/4$ for the A1(fcc) crystal, and $v_0 = a_0^3/8$ for the other solids in this study.

For the present study we use the structureless pseudopotential model[54] or "stabilized jellium" EOS (SJEOS).[55] It is almost ideally suited for the description of the regime close to the equilibrium volume. The form of the SJEOS is motivated by a physical picture of cohesion. We fit the SJEOS to the energy-volume data by minimizing the least-square error. As a check, we also used the Murnaghan EOS, which is more standard but has no microscopic foundation. In the present paper, we give $B_0$ in units of GPa (1 a.u. = 29421 GPa).

## IV. LATTICE CONSTANTS

The experimental lattice constants include zero-point phonon effects (ZPPEs), and are often measured at room temperature. These experimental values are not directly comparable with the results of ground-state electronic structure calculations (0 K). We show here that neglecting these frequently overlooked effects might invalidate any comparison of experiment and theory. The experimental low temperature (5-50 K) lattice constants values are from Ref. 56 (Li, Sr), Ref. 57 (Na, K, Rb, Cs, and Ba), Ref. 58 (Ca, Al, Pb, Cu, Rh, Pd, and Ag), and Ref. 59 (NaCl). The rest are based on room temperature values from Ref. 60 (C, Si, SiC, Ge, GaAs NaF, LiF, MgO), and from Ref. 57 (LiCl), corrected to the $T = 0$ limit using thermal expansion corrections from Ref. 58. For MgO Ref. 61 give a lattice constant at 77 K that is smaller than our estimated 0 K data and we use that value (4.203 Å). (Note that a linear extrapolation of the lattice constant from 300K to 0K is neither accurate nor used here.) For lattice constants the ZPPEs manifest as zero-point anharmonic expansion (ZPAE). This effect may be estimated from Eq. (A6) of Ref. 55

$$\frac{\Delta a_0}{a_0} = \frac{\Delta v_0}{3 \cdot v_0} = \frac{3}{16}(B_1 - 1)\frac{k_B \cdot \Theta_D}{B_0 \cdot v_0}. \qquad (7)$$

The ZPAE was estimated from experimental $a_0$, $B_0$, $v_0$, and $\Theta_D$ (Debye temperature) and from corrected theoretical SJEOS $B_1^{TPSS}$ values as described in Ref. 38. Note that $v_0$ in Eq (7) is the volume per atom in the crystal. The ZPAE corrections for C, Si, SiC, Ge, GaAs, NaCl, NaF, LiCl, LiF, and MgO are incorrectly given in Ref. 38; those values should be multiplied by 2 (as in the errata to Refs. 14 and 38). The magnitude of this correction is in the range of 0.003 and 0.046 Å (LiF), and it is relatively large for alkali metals (0.8, 0.4, and 0.3 % for Li, Na, and K, respectively) and ionic solids (1.2 and 0.5 % for LiF and NaCl, respectively). The neglect of this effect can be justified for benchmarking LDA and PBE functionals, where the average errors (-1.3 % and +1.6%, respectively, cf. Table I) are considerably larger than the errors arising from the neglect of ZPAE. As the zero-point anharmonic motion always expands the lattice, the neglect of it introduces a systematic bias in the appraisal of the functionals. The average expansion is +0.015 Å (~0.35 %) for the ZPAE values shown in Table II. The uncorrected experimental results are closer to PBE than LDA, while the ZPAE-corrected experimental values are smaller and thus move closer to LDA values. Table I also shows that recent functionals developed for solids are considerably closer to the corrected experimental values; the mean errors (ME) in Table I are about 0.01 Å ( ~0.25 %) for the PBEsol and second-order generalized gradient approximation (SOGGA)[15] functionals. Hence, neglecting the ZPAE biases the rating of such functionals. Note that in the original SOGGA paper[15] the ZPAE corrections for non metallic solids are incorrectly given. Consequently our statistics for SOGGA in the Table I. is different from the published statistics,[15] and the agreement between SOGGA and the experiment is slightly worse here.

The good agreement between PBEsol and SOGGA is particularly interesting since the SOGGA exchange enhancement factor was constructed from half-and-half mixing of PBE and RPBE[62], using the exact gradient expansion coefficient ($\mu = 10/81$) in the same way as suggested for PBEsol. The main difference between the exchange enhancement factors is that SOGGA enforces a smaller value for the large gradient limit (tighter Lieb–Oxford bound), 1.552 instead of 1.804 used in PBE and PBEsol. Fig. 1 shows that the SOGGA and PBEsol $F_x(s)$ curves are very close to each other for small gradients ($s < 2$). The SOGGA functional uses the unchanged PBE correlation functional. Consequently the origin of the PBEsol improvement over PBE in lattice constants for solids is to be found in the modification of the exchange functional. This is in agreement with the explanation given in the original PBEsol paper.[14]

The AM05 exchange functional[19] is based on the Airy gas paradigm of Kohn and Mattsson.[17] For small $s$, the AM05 and PBEsol exchange functionals are quite different (cf. Fig S1 of Ref. 14): the PBEsol $F_x(s)$ follows the exact gradient expansion while the AM05 $F_x(s)$ remains close to 1 if $s < 1$). Figure 1 shows that for larger $s$ the AM05 gradient enhancement factor increases rapidly, crosses the PBE curve, and finally rises at $s \approx 5.2$ above 1.804, the maximum value that ensures satisfaction of the Lieb-Oxford bound for all possible densities. Indeed, AM05 will violate this lower bound on the exchange energy for any density in which $s$ is sufficiently large everywhere, which can be achieved by starting with a suitable density $n(\mathbf{r})$ for $N \gg 1$ electrons and scaling it down to $n(\mathbf{r})/N$ so that $s \sim N^{1/3}$ becomes everywhere greater than 5.2. For example, the density of an atom has $s > 0$ everywhere. Note that the LAG functional[18] behaves very similarly. Despite the difference in the exchange enhancement functions (cf. Fig. 1), the AM05 and PBEsol functionals have very similar exchange-correlation enhancement factors $F_{xc}(s,r_s,\zeta=0)$ for $r_s =1$ as demonstrated in Fig.. 2 for the spin-unpolarized density, where $r_s$ is the Wigner-Seitz radius and $\zeta = (n_\uparrow - n_\downarrow)/n$ is the relative spin polarization. For small $s$, the two functionals are quite similar for electron densities around $1 < r_s < 5$ (typical for valence and core-valence regions). However, for large $s$ and $r_s$ values, the two functionals behave quite differently. Fig. 2 shows that the AM05 curves go below the PBEsol curves for $s <1$, and after going through a minimum they cross the PBEsol curves and increase more steeply than the PBEsol curves for large $s$. The difference between the AM05 and PBEsol curves increases with $s$ (and with $r_s$ for $r_s >2$).

Inspection of the PBE results in Table I shows that the BAND/LCAO and VASP/PAW results agree quite well with each other. The calculated lattice constants depend on the choice of the Gaussian basis set. Application of the GTO1 or GTO2 basis sets gives 0.010-0.012 Å longer lattice constants on average, compared to converged BAND/LCAO and VASP/PAW results. Consequently non-converged Gaussian basis sets might slightly bias the estimation of the performance of the functionals that reach a high accuracy like PBEsol and SOGGA (cf. mean error ME $\cong$ 0.010 Å, and mean absolute error MAE $\cong$ 0.023 Å in Table I). Even the estimation of the performance of the TPSS in Ref. 13 might be biased by 0.01 Å. The TPSS BAND/LCAO results agree better with experiment than TPSS/GTO1 results (cf. MEs = 0.048 and 0.058 Å, and MAEs = 0.052 and 0.059 Å, respectively; for the BAND results see Table I). Comparison with the VASP or BAND results show that the more expensive GTO2 basis set is somewhat better for diamond and Si than the GTO1 basis set, but no clear improvement can be observed for SiC, Ge, GaAs, and MgO. An earlier BAND study shows that the inclusion of scalar relativistic effects shortens the lattice constants of Cu (-1%) and Ag (-2.4%).[51] The all electron (non-relativistic) GTO calculations agree well with the relativistic BAND results for Cu due to the GTO1 basis set error. The relativistic effective core potential (ECP) basis sets used for Rh, Pd, and Ag in Refs 14 and 38 give mixed results: a good agreement with BAND for Pd, shorter lattice constant for Ag (-0.02 Å), and longer for Rh (+0.04 Å).

The mean relative error (MRE) of the PBEsol results in Table I (MRE $\cong$ 0.2 %) lies between those of PBE (MRE $\cong$ 1.4 %) and LDA (MRE $\cong$ -1.3 %). The SOGGA/GTO1 results are also excellent, but a small GTO1 basis set error is included in these results. Removing this small error (e.g., by BAND) will not deteriorate the SOGGA statistics for these 18 solids. Note that in the evaluation of the SOGGA functional[15] two small errors, the basis set error and the 10 incorrect experimental references compensate each other. The TPSS BAND/LCAO results (MRE $\cong$ 1.0 %) in Table I show some improvement compared to PBE results, but they do not reach the quality of the PBEsol results.

The recent AM05 functional performs well too, but the lattice constants are slightly too long on average and thus less accurate compared to PBEsol or SOGGA results. Fig. 3 shows the individual relative errors (%) of the lattice constants calculated with PBEsol, AM05, and SOGGA compared to the ZPAE-corrected experimental lattice constants at 0 K. The larger errors of the AM05 functional for bulk K, NaCl and NaF contribute to the larger statistical error of the AM05 functional for this test set comprising 18 solids. The AM05 lattice constants are VASP/PAW values taken from Ref. 48, except for K and Ge, which were computed for this work. According to the mean absolute relative errors (MARE) in Table I, the order of accuracy is PBEsol $\cong$ SOGGA > AM05 > TPSS > PBE $\cong$ LDA.

The systematic deviation between AM05 and PBEsol and the performance of the TPSS functional will be discussed on a larger test set, in support of statements made by some of us in Ref. 32. Table II shows the Strukturbericht symbols, the LDA, PBEsol, and PBE lattice constants for 24 solids calculated with the BAND/LCAO program, together with the ZPAE-corrected experimental values (Å) and the ZPAE corrections (Å). Comparison of PBEsol/BAND results with the EMTO results[22] shows that this latter method results in too long lattice constants for Cs, Ca, and Sr; the difference is 0.01-0.02 Å. Detailed comparison shows that the difference between AM05 and PBEsol lattice constants increases systematically with increasing atomic number for alkali and alkaline earth metals. For Rb and Cs the AM05 lattice constants are quite far from the PBEsol results (larger by 0.10 and 0.15 Å, respectively), and agree well with the PBE lattice constants.[22] Similar effects can be observed for Sr and Ba bulk metals.[22] As discussed earlier (cf. Fig. 3), PBEsol outperforms AM05 for ionic insulators and heavier alkali metals while AM05 outperforms PBEsol for alkaline earth metals. In this sense the recent observation[31] that AM05 and PBEsol yield identical results for a wide range of solids is not valid for a large number of solids like heavier alkali and alkaline earth metals and ionic insulators.

Inspection of the results in Table II shows that the TPSS functional performs well for Li, Na, Ca, Sr, Ba, Al, Cu, and Rh, and it gives too long lattice constants for K, Rb and Cs, making the TPSS results worse than PBE for these latter metals. Interestingly, for Ca, Sr and Ba the TPSS results are quite good (MARE = 0.3%) and agree better with the PBE (MARE = 0.4%) and AM05 (MARE = 1%) results than with the too short PBEsol results (MARE = 2%), while for Al, Cu and Rh the TPSS results (MARE = 0.3%) agree well with the PBEsol results (MARE = 0.3%). The AM05, and PBE lattice constants are too long for these solids; MARE = 0.5, and 0.9%, respectively. This shows the potential of the meta-GGA to alter trends, as for both groups of solids its MARE remains around 0.3%.

Table III shows the maximal values of the reduced exchange gradient $s$ calculated for our solids from BAND electron densities. The values show that Li, the ionic solids, Ge and GaAs have the largest maximal $s$ (2.1-2.2), while this value is considerably smaller for the other solids (0.8 < $s$ < 1.7). This is the explanation for the surprising similar performance of the AM05 and PBEsol functionals for many metals, and the larger differences for ionic solids, where the large $s$ and $r_s$ region is more important than in metals.

Fuchs, Bockstedte, Pehlke and Scheffler[63] presented convincing evidence that two density functionals that reduce to LDA for a uniform density can produce different lattice constants largely through their differences in the region of core-valence overlap, and not in the pure valence region. This conclusion also seems supported by the analysis of Ruban and Abrikosov.[64] Fig. 2 of Ref. 65 plots $s$ and $r_s$ vs. distance from the nucleus for the nitrogen atom, showing that $r_s \approx 1$ or less and 0.3 < $s$ < 1.3 in the core-valence overlap region. Our Fig, 2 shows an especially close agreement between AM05 and PBEsol for all $s$ at $r_s \approx 1$, and reasonable agreement for $s < 1$ at $r_s$ away from 0. These features might explain the rough agreement of AM05 and PBEsol lattice constants for most solids, and their close agreement for solids with $s < 1$ everywhere. We suspect that the maximum $s$ values for solids in our Table III tend to occur in the core-valence overlap region. These issues deserve further study.

The results in Table IV show that the performance of the functionals is different for metallic and non-metallic solids compared to thermally- and ZPAE-corrected experimental results. Note that Refs. 22 and 41 use partially or uncorrected room temperature experimental lattice constants as reference values. In Table IV we also show the performance of the same functionals compared to the same experimental values used in Refs 22 and 41.

For the 14 metals in this test set, the PBEsol functional is the best performer, giving slightly shorter lattice constants than the fully corrected experimental values (MRE = -0.7%, MARE = 0.8%). PBE performs quite well and gives slightly too long lattice constants by about 1% (MARE = 1.24%). The LDA MRE is the largest, -2.7%. Our results for the main group metals can be compared to the results in Table II of Ropo et al.[22] for the same metals. As noted earlier, the calculated EMTO lattice constants are in reasonably good agreement for LDA, PBEsol and PBE with our calculated lattice constants in Table II. However, for this test set Ropo et al. conclude that PBE is the best performer. The origin of the different conclusion is the neglect of the ZPAE for all metals and the use of room temperature experimental lattice constants for Al, Pb, Cu, Rh, Pd and Ag (cf. the good agreement between PBE and partially or uncorrected room temperature experimental results in Table IV). According to the MAREs for corrected experimental lattice constants in Table IV, the order of accuracy is PBEsol > TPSS > PBE >> LDA for metals.

For the 10 non-metals in this test set, the PBEsol and LDA perform almost equally well, giving opposite +0.6 and −0.7 % MREs, respectively, while PBE gives MRE $\cong$ +1.8%. The LDA performs very well for the lattice constants of our non-metallic solids, but studies that ignore the ZPAE might easily draw wrong conclusions. The results in Table IV show that ignoring ZPAE effects shifts the MRE by 0.52% (the ME by about 0.02 Å) away from the LDA. Note that PBEsol gives the same but opposite error for metals and non-metals, and this contributes to its good performance for the whole test set (cf. Tables II and IV). According to the MAREs for corrected experimental lattice constants in Table IV, the order of accuracy is PBEsol > LDA > TPSS > PBE for non-metals.

## V. BULK MODULI

Temperature and phonon effects can modify the bulk modulus up to 20% for Li[66] and 5-8% for the other metals studied here. The temperature effects are about 5-15%, the ZPPEs are about 2% on average (span 1-5% range). The experimental error is up to 5-10%.

Table V shows the calculated and experimental[38,55,60] bulk moduli (GPa) for our extended set of 24 solids. The experimental references used for Table V are corrected for zero temperature but do not include ZPPEs. However, for main group metals, the estimated ZPPEs are given in Ref. 55 (cf. Table V and eq. S12), and we show them in %. The ZPPEs make the corrected experimental bulk moduli 2% stiffer on average. As bulk moduli of these solids span a large, more than two-order-of-magnitude (2-440 GPa) range, we also show the mean relative error (MRE %) and the mean absolute relative error (MARE %). Investigation of the results in Table V shows that the LDA overestimation of the bulk moduli is about 15% (too stiff) and the PBE underestimation is about 9% (too soft, but considerable improvement over LDA). The PBEsol performance is excellent, giving about a 1% overestimation. The relevant AM05/VASP results[48] are on the PBE side (too soft) and quite close to the experimental results, while PBEsol is on the LDA side and again close to the experimental results. The ZPPE corrections would shift the corrected experimental values in the direction of LDA and PBEsol and worsen the agreement between experiment and AM05 or PBE results. Gaudoin and Foulkes[66] give $B_0$ values after removal of finite temperature and zero-point effects for Li, Al, and Pb: 14.5, 81.3, and 47.3 GPa, respectively. These values agree well with our PBEsol values in Table V.

Using GTO1 basis set makes the solids on average stiffer by 2% for LDA and PBE compared to BAND or earlier VASP results.[48] Thus the GTO1 results in Ref. 38 might contain a 2-3% random basis set error. The 0.3% average deviation between PBEsol/BAND and VASP bulk moduli shows the precision of our current calculations. The BAND and VASP (not shown) agree well with each other for LDA, PBE and PBEsol bulk moduli. We plan to include more $3d$, $4d$, and $5d$ metals, and other nonmetallic solids in our test set. However, the lack of good experimental bulk moduli at 0 K limits our effort to expand the test set.

The large mean absolute relative errors of the bulk modulus (15% for LDA, 9% for PBE) shown in Table V are reduced to 5% by PBEsol. One can alternatively achieve this level of accuracy by combining LDA or PBE equation-of-state parameters with the experimental value of the lattice constant; see Eq. (21) and Table IV of Ref. 38.

## VI. COHESIVE ENERGIES

Table VI lists the cohesive energies (eV/atom) obtained for 18 solids from PBE and PBEsol functionals using different methods (GTO, VASP/PAW and BAND/LCAO). In GAUSSIAN, the basis functions used to describe core electrons must be the same in the solid and in the atom, so that basis-set limitations in the core will cancel out of the cohesive energy. But converged energies for the valence electrons in a solid can be achieved without the more diffuse basis functions needed to converge the energy of the valence electrons in a free atom or molecule. These more diffuse basis functions can create computational problems[38] for our solids other than C, Si, SiC, Ge, and GaAs. With this in mind, we have calculated cohesive energies from GAUSSIAN for those other solids, using different GTO basis functions for the atom (standard molecular basis sets) and for the valence electrons in the solid (standard[38,41] pruned versions thereof), and these cohesive energies are largely confirmed by those from our BAND and VASP calculations.

Ref. 38, using the *same* GTO basis sets for the solid as for the free atom, was only able to report four cohesive energies (C, Si, SiC, Ge). Ref. 38 was able to report four more cohesive energies (NaCl, NaF, LiCl, LiF), using different basis sets for the cation in the solid and for the corresponding free atom (as we do), but questioned whether this could work for the metals. It was found that PBE performs considerably better than LDA (serious overbinding) or TPSS (underbinding), but the small test set of 8 non-metals was insufficient for establishing trends.[38] A more complete study[42] using VASP on a test set different from but overlapping with our test set (with only K and Ge missing) shows that PBE is better for cohesive energies than HSE[67] or PBEh.[42] In Table VI we show the relevant PBE VASP results from Ref. 42. Comparison of GTO2 and VASP results shows a good agreement, except for Cu and Rh. Note that the GTO2 result for Cu is non-relativistic. Similar agreement between GTO2 and BAND results is shown in Table VI for the PBEsol functional. Comparison of PBEsol results in Table VI and Ref. 15 shows relatively large disagreement for ionic solids (more than 0.1 eV).

We collected the 0 K experimental results from Ref. 68 with the experimental errors where available. The experimental cohesive energies were corrected by the zero-point vibration energy (ZPVE)[51] calculated from the Debye temperature $\Theta_D$,

$$ZPVE = (9/8)k_B\Theta_D. \qquad (8)$$

The values in Table VI show that the ZPVE is frequently comparable in magnitude to the experimental error. The results in Table VI show the general overbinding tendency of PBEsol. PBE performs better than PBEsol except for ionic solids where PBEsol shows an excellent performance.

Note that evaluation of the cohesive energy requires a generalization of the density functional to a spin-density functional ($\zeta > 0$), which was published for most functionals but not for AM05 at the time this paper was written; however, see Refs. 69, 70. BAND and VASP have a spherical, spin-unpolarized reference atom, but we have used the real atom in our cohesive energy calculations. The atomic corrections, which turn the energy of the reference atom into that of the real one, are available from the authors.

## VII. CONCLUSIONS

We have shown that neglecting the thermal and zero-point phonon effects might invalidate any comparison of experiment and theory for lattice constants and bulk moduli. The uncorrected experimental results are much closer to PBE than LDA. For 24 solids in our test set, after correction of the experimental data, PBE systematically overestimates the lattice constants (by 1.3%) and underestimates the bulk moduli (by 9%), while the LDA results show larger and opposite errors (1.9 and 15%, respectively). Mean errors (ME) of recent functionals developed for solids like PBEsol are about 0.01 Å (~0.25 %) for the lattice constants and 1% for the bulk moduli. Hence, neglecting the effect of the zero-point anharmonic expansion (ZPAE), +0.015 Å (~0.35 %), biases the judgment about the performance of such functionals. Neglecting the thermal expansion adds further bias (up to 1.4 %).

For the lattice constants of our 10 non-metals, PBEsol and LDA perform almost equally well, giving opposite +0.6 and −0.7 % average relative errors, respectively. PBEsol gives the similar but opposite average relative error for metals (-0.7%) and non-metals (0.6%); this contributes to its good performance for the whole test set of 24 solids. The PBE functional shows poor performance for non metals and quite good performance for metals where LDA fails.

The SOGGA functional uses a PBEsol-like exchange functional for $s < 2$ and the PBE correlation functional (not fitted to the surface exchange-correlation energy of jellium, unlike the correlation functionals of AM05 and PBEsol). The PBEsol and SOGGA lattice constants agree quite well. Consequently the origin of the PBEsol improved lattice constants for solids is to be found in the modification of the exchange functional. This is in agreement with the explanation given in our original PBEsol paper.[14]

In section IV, we have proposed an explanation, in terms of the exchange-correlation enhancement factor $F_{xc}(s,r_s)$, for the close similarity of AM05 and PBEsol lattice constants in solids with $s < 1$ everywhere, and the greater difference for some solids with $s_{max} \gg 1$. Our explanation is consistent with the importance of exchange-correlation nonlocality in the core-valence overlap region.

The Gaussian basis sets introduce a small 0.005-0.009 Å (0.2%) lattice constant lengthening that slightly biases the assessment of the functionals, but does not change the conclusion. Our previous conclusions based on Gaussian basis sets remain valid and supported by VASP and BAND results. This shows that carefully selected Gaussian basis sets might be suitable for testing density functionals, despite the serious problems of basis-set construction.

The Gaussian basis sets introduce 2-3% uncertainty into the calculated bulk moduli, while the VASP and BAND results agree within 0.3%. These errors are negligible compared to the experimental errors (up to 10%) and the errors arising from neglect of thermal (up to 15%) and zero-point phonon effects (1-3%, up to 4.5%).

For cohesive energies of the 18 solids, PBEsol shows an overbinding tendency (by 0.22 eV/atom on average). PBE slightly underbinds (by 0.13 eV/atom on average) and performs better than PBEsol except for alkali metals and ionic solids where PBEsol shows an excellent performance.

The results suggest that possibly no single GGA can describe with high accuracy the surface energies, lattice constants, bulk moduli, and cohesive energies of solids at the same time. The original PBE is biased toward a correct description of atoms and molecules, while PBEsol is biased toward solids. Many GGA variants that do not build on the exact gradient expansion for exchange might give accurate lattice constants. Restoring the gradient expansion for exchange over a wide range of reduced density gradients (as in PBEsol) might not be necessary for good lattice constants for a limited class of solids, but is needed to construct more universal approximations.[32] The TPSS meta-GGA provides an excellent description of atomic total energies, molecular atomization energies, and jellium surface energies, but its lattice constants might be improved by imposing this new condition.

In short, the PBEsol GGA for solids works well for the lattice constants and bulk moduli of typical nonmolecular solids. An accurate lattice constant and bulk modulus may[4] be accompanied by a good description of thermal effects. For the open-shell 3$d$ transition metals, PBE is better[22,24] than PBEsol, but these solids are bonded in part by the highly localized 3$d$ orbitals to which the second-order gradient expansion of the exchange energy (on which PBEsol is based) may not apply.[22] Under sufficiently intense compression, all solids (including the 3$d$ metals) should be better described[14] by PBEsol.

Since PBE is much better than PBEsol for the total energies of atoms, and for the atomization energies of molecules, we expected that PBE would also be better for the cohesive energies of solids. While this has been confirmed here in a statistical sense, we also find unexplained special cases (the alkali metals and the alkali halides, where the atoms have one electron outside a closed shell or one electron missing from a closed shell) where PBEsol cohesive energies are excellent and much better than PBE. For a functional that will be accurate over a much wider range of systems, we intend to look beyond the GGA level to an improved meta-GGA. A meta-GGA form is more flexible, and computationally not much slower than a GGA, making it the natural successor of LDA and GGA in applications.

NOTE ADDED IN PROOF: After this article was accepted, we learned of another[71] lattice-constant test of semilocal functionals, for 60 solids using the WIEN2K code, with results similar to ours. The mean absolute deviation in Angstrom of each column of our Table II (excluding Cs, not studied in Ref. 71) from the results of Ref. 71 is 0.005 (LDA), 0.008 (PBEsol, AM05), 0.009 (PBE), 0.012 (TPSS), and 0.003 (Expt.-ZPAE). The WIEN2K SOGGA results of Ref. 71 are also somewhat different from the GTO1 results published in Ref. 15 for 18 solids (with deviations due to the Gaussian basis set error in the range of -0.021 to +0.035 Å, and mean absolute deviation 0.015 Å), but the overall statistics for SOGGA lattice constants remains good, close to the PBEsol statistics as we have predicted in this paper.

**ACKNOWLEDGMENTS**

J.P.P. thanks Fabien Tran for pointing out the errors of the ZPAE calculated from Eq. (7) for C, Si, SiC, and the alkali halides in Refs. 14 and 38. J.P.P. thanks Levente Vitos for pointing out Ref. 64. J.P.P., A. R. and G.I.C thank NSF (Grant No. DMR-0501588) and MTA-NSF (travel grant) for support. S.L. acknowledges financial support from ANR PNANO Grant No. ANR-06-NANO-053–02; J.G.A. and S.L. acknowledge financial support from ANR Grant No. ANR-BLAN07–1-186138.

TABLE I. Statistical data for the equilibrium lattice constants (Å) of the 18 test solids of Ref. 38 at 0 K calculated from the SJEOS. The Murnaghan EOS yields identical results within the reported number of decimal places. Experimental low temperature (5-50 K) lattice constants are from Ref. 56 (Li), Ref. 57 (Na, K), Ref. 58 (Al, Cu, Rh, Pd, Ag), and Ref. 59 (NaCl). The rest are based on room temperature values from Ref. 60 (C, Si, SiC, Ge, GaAs, NaF, LiF, MgO) and Ref. 57 (LiCl), corrected to the $T = 0$ limit using the thermal expansion from Ref. 58. An estimate of the zero-point anharmonic expansion (ZPAE) has been subtracted out from the experimental values (c.f. Table II). (The calculated values are precise to within 0.001 Å for the given basis sets, although Gaussian GTO1 and GTO2 basis-set incompleteness limits the accuracy to 0.02 Å.) GTO1: The basis set used in Ref. 38. GTO2: For C, Si, SiC, Ge, GaAs, and MgO, the basis sets were taken from Ref. 41. For the rest of the solids, the GTO1 basis sets and effective core potentials from Ref. 38 were used. The best theoretical values are in boldface. The LDA, PBEsol, and PBE GTO2 results are from Ref. 14. The SOGGA GTO1 results are from Ref. 15.

|  | LDA | LDA | PBEsol | PBEsol | PBEsol | AM05 | SOGGA | PBE | PBE | PBE | TPSS |
|---|---|---|---|---|---|---|---|---|---|---|---|
|  | GTO2 | VASP | GTO2 | BAND | VASP | VASP | GTO1 | GTO2 | VASP | BAND | BAND |
| ME[a] (Å) | -0.047 | -0.055 | 0.022 | **0.010** | 0.012 | 0.029 | **0.009** | 0.075 | 0.066 | 0.063 | 0.048 |
| MAE[b] (Å) | 0.050 | 0.050 | 0.030 | **0.023** | 0.023 | 0.036 | **0.024** | 0.076 | 0.069 | 0.067 | 0.052 |
| MRE[c] (%) | -1.07 | -1.29 | 0.45 | **0.19** | 0.24 | 0.58 | **0.19** | 1.62 | 1.42 | 1.35 | 0.99 |
| MARE[d] (%) | 1.10 | 1.15 | 0.67 | **0.52** | 0.52 | 0.80 | **0.50** | 1.65 | 1.48 | 1.45 | 1.10 |

[a]Mean error.
[b]Mean absolute error.
[c]Mean relative error; (calculated-experimental)/experimental 100%.
[d]Mean absolute relative error.

Table II. Strukturbericht symbols (Str.) and equilibrium lattice constants (Å) of 24 test solids calculated with BAND/LCAO from the SJEOS. The Strukturbericht symbols are used for the structure as follows: A1, fcc; A2, bcc; A4, diamond; B1, rock salt; B3, zinc blende. The Murnaghan EOS yields identical results within the reported number of decimal places. Low temperature (5-50 K) experimental lattice constants values are from Ref. 56 (Li, Sr), Ref. 57 (Na, K, Rb, Cs, and Ba), Ref. 58 (Ca, Al, Cu, Rh, Pd, and Ag), and Ref. 59 (NaCl). The rest are based on room temperature values from Ref. 60 (C, Si, SiC, Ge, GaAs NaF, LiF, MgO) and Ref. 57 (LiCl), corrected to the $T = 0$ limit using the thermal expansion from Ref. 58. An estimate of the zero-point anharmonic expansion (ZPAE) is subtracted out from the experimental values (shown in boldface). The best theoretical values are also in boldface. We show, for reference, the AM05 values from VASP. For K, Ge, Rb, Cs, Ca, Sr, Ba, and Pb, we have computed the AM05 values for this work; for the other solids, we have taken the AM05 values from Ref. 48. Note that, for the alkali and alkaline-earth metals and alkali halides, AM05 values are often closer[32] to PBE than to PBEsol.

| Solid | Str. | LDA | PBEsol | AM05 | PBE | TPSS | Expt.-ZPAE | ZPAE |
|---|---|---|---|---|---|---|---|---|
| Li | A2 | 3.363 | 3.428 | **3.455** | 3.429 | **3.445** | **3.449** | 0.028 |
| Na | A2 | 4.054 | 4.167 | **4.212** | 4.203 | 4.240 | **4.210** | 0.015 |
| K | A2 | 5.046 | **5.210** | 5.297 | 5.284 | 5.360 | **5.212** | 0.013 |
| Rb | A2 | 5.373 | **5.561** | 5.670 | 5.667 | 5.736 | **5.576** | 0.009 |
| Cs | A2 | 5.751 | **5.991** | 6.182 | 6.207 | 6.241 | **6.039** | 0.006 |
| Ca | A1 | 5.328 | 5.446 | 5.474 | **5.521** | 5.524 | 5.553 | 0.011 |
| Sr | A1 | 5.782 | 5.901 | 5.966 | **6.004** | 5.988 | **6.045** | 0.008 |
| Ba | A2 | 4.747 | 4.866 | 4.957 | **5.022** | 4.973 | 4.995 | 0.005 |
| Al | A1 | 3.985 | **4.013** | 4.004 | 4.037 | **4.009** | **4.020** | 0.012 |
| Pb | A1 | 4.874 | **4.926** | 4.939 | 5.035 | 4.984 | **4.902** | 0.003 |
| Cu | A1 | 3.517 | 3.562 | 3.565 | 3.628 | **3.575** | **3.595** | 0.007 |
| Rh | A1 | 3.755 | **3.780** | 3.773 | 3.829 | **3.803** | **3.793** | 0.005 |
| Pd | A1 | 3.836 | **3.876** | 3.872 | 3.942 | 3.903 | **3.875** | 0.004 |
| Ag | A1 | 4.010 | **4.053** | 4.054 | 4.147 | 4.086 | **4.056** | 0.005 |
| C | A4 | **3.532** | 3.553 | 3.551 | 3.569 | 3.568 | **3.543** | 0.023 |
| Si | A4 | **5.403** | 5.431 | 5.431 | 5.466 | 5.451 | **5.416** | 0.014 |
| SiC | B3 | **4.329** | 4.356 | 4.350 | 4.377 | 4.366 | **4.342** | 0.018 |
| Ge | A4 | **5.623** | 5.675 | 5.678 | 5.759 | 5.721 | **5.640** | 0.012 |
| GaAs | B3 | **5.605** | 5.661 | 5.672 | 5.746 | 5.713 | **5.638** | 0.010 |
| NaCl | B1 | 5.465 | **5.602** | 5.686 | 5.700 | 5.703 | **5.565** | 0.029 |
| NaF | B1 | **4.502** | 4.629 | 4.686 | 4.705 | 4.705 | **4.579** | 0.030 |
| LiCl | B1 | 4.968 | **5.058** | 5.119 | 5.142 | 5.094 | **5.056** | 0.032 |
| LiF | B1 | **3.913** | 4.003 | 4.039 | 4.062 | 4.027 | **3.964** | 0.046 |
| MgO | B1 | **4.168** | 4.223 | 4.232 | 4.255 | 4.237 | **4.184** | 0.019 |

TABLE III. Maximal values of the reduced gradient, $s$ (a. u.) in various solids calculated by BAND/LCAO. A region of radius 0.2 bohr/Z around the nucleus, where relativistic effects might be important, has been excluded.

| Solid | Max. $s$ |
|---|---|
| Li | 2.1 |
| Na | 1.9 |
| K | 1.7 |
| Rb | 1.6 |
| Cs | 1.5 |
| Ca | 1.3 |
| Sr | 1.3 |
| Ba | 1.1 |
| Al | 1.4 |
| Pb[a] | 0.8 |
| Cu | 1.0 |
| Rh | 0.8 |
| Pd | 0.8 |
| Ag | 0.9 |
| C | 1.4 |
| Si | 1.4 |
| SiC | 1.5 |
| Ge | 2.2 |
| GaAs | 2.2 |
| NaCl | 2.1 |
| NaF | 2.1 |
| LiCl | 2.2 |
| LiF | 1.9 |
| MgO | 1.6 |

[a] Largest significant value (the integration weight is larger than $10^{-3}$).

TABLE IV. Statistical data, mean error (ME), mean absolute error (MAE), mean relative error (MRE %), and mean absolute relative error (MARE %), for lattice constants (Å) of the 14 metals and 10 non-metals in the test set of 24 solids calculated with BAND/LCAO from the SJEOS. Comparisons to thermally- and ZPAE-corrected experimental results (left) and to partially or uncorrected room temperature experimental values used in Refs 22 and 41 (right). The best agreements with the experiment are in boldface. For the AM05 values of Table II, compared to corrected experimental results, the total ME and MAE are 0.025 Å and 0.048 Å, respectively. The AM05 functional performs better for metals (MAE = 0.045 Å) than for non-metals (MAE = 0.052 Å).

| Solid | Compared to corrected experimental values | | | | Compared to experimental values used in Refs 22 and 41 | | | |
|---|---|---|---|---|---|---|---|---|
| | LDA | PBEsol | PBE | TPSS | LDA | PBEsol | PBE | TPSS |
| Metals (14) | | | | | | | | |
| ME (Å) | -0.136 | **-0.039** | 0.046 | **0.039** | -0.151 | -0.054 | **0.030** | **0.024** |
| MAE (Å) | 0.136 | **0.042** | 0.060 | 0.060 | 0.151 | **0.058** | 0.055 | 0.060 |
| MRE (%) | -2.71 | **-0.76** | 0.95 | **0.74** | -3.04 | -1.10 | **0.61** | 0.39 |
| MARE[a] (%) | 2.71 | **0.83** | 1.24 | 1.15 | 3.04 | **1.21** | **1.15** | **1.19** |
| Non-metals (10) | | | | | | | | |
| ME (Å) | -0.042 | **0.026** | 0.085 | 0.066 | -0.067 | **0.001** | 0.060 | 0.040 |
| MAE (Å) | 0.042 | **0.026** | 0.085 | 0.066 | 0.067 | **0.001** | 0.060 | 0.043 |
| MRE (%) | -0.86 | **0.56** | 1.76 | 1.35 | -1.41 | **0.00** | 1.19 | 0.79 |
| MARE[a] (%) | 0.86 | **0.56** | 1.76 | 1.35 | 1.41 | **0.31** | 1.19 | 0.84 |
| Total (24) | | | | | | | | |
| ME (Å) | -0.097 | **-0.012** | 0.062 | 0.050 | -0.116 | **-0.031** | 0.043 | 0.031 |
| MAE (Å) | 0.097 | **0.036** | 0.070 | 0.062 | 0.116 | **0.040** | 0.057 | 0.053 |
| MRE (%) | -1.94 | **-0.21** | 1.29 | 0.99 | -2.36 | **-0.64** | 0.85 | **0.56** |
| MARE[a] (%) | 1.94 | **0.72** | 1.45 | 1.23 | 2.36 | **0.84** | 1.17 | 1.04 |

[a] (calculated-experimental)/experimental 100%.

Table V. Bulk moduli (GPa), zero-point phonon effects (ZPPE %), and statistical data of 24 test solids calculated with BAND/LCAO from the SJEOS. The experimental data include thermal corrections but not the ZPPE, as ZPPE values are not available for most of the solids. The available ZPPE corrections are taken from Ref. 55. The best theoretical values are in boldface.

| Solid | LDA | PBEsol | PBE | Expt. | ZPPE (%) |
|---|---|---|---|---|---|
| Li | 15.2 | **13.8** | **13.8** | 13.3 | 4.5 |
| Na | 9.50 | 8.16 | **7.86** | 7.50 | 3.1 |
| K | 4.60 | **3.74** | 3.44 | 3.70 | 2.1 |
| Rb | 3.54 | **2.95** | 2.76 | 2.90 | 1.4 |
| Cs | 2.58 | **1.96** | 1.72 | 2.10 | 0.9 |
| Ca | 19.1 | **17.8** | 17.0 | 18.4 | 1.8 |
| Sr | 14.8 | 13.2 | **12.1** | 12.4 | 1.3 |
| Ba | 10.9 | **9.06** | 7.91 | 9.30 | 0.8 |
| Al | 83.8 | 82.6 | **78.0** | 79.4 | 3.3 |
| Pb | 54.3 | **48.1** | 37.1 | 46.8 | 1.1 |
| Cu | 190 | 166 | **142** | 142 | |
| Rh | 320 | 296 | **260** | 269 | |
| Pd | 227 | **205** | 169 | 195 | |
| Ag | 139 | **119** | 91 | 109 | |
| C | 467 | 450 | **434** | 443 | |
| Si | **96.8** | 94.2 | 89.2 | 99.2 | |
| SiC | **225** | 218 | 210 | 225 | |
| Ge | **72.6** | 68.1 | 59.7 | 75.8 | |
| GaAs | **74.2** | 69.1 | 61.3 | 75.6 | |
| NaCl | 32.4 | **25.8** | 23.6 | 26.6 | |
| NaF | 61.2 | **48.6** | 44.4 | 51.4 | |
| LiCl | 40.8 | **35.2** | 32.1 | 35.4 | |
| LiF | 86.5 | **73.1** | 67.1 | 69.8 | |
| MgO | **172** | 157 | 149 | 165 | |
| | | | | | |
| ME[a] (GPa) | 10.2 | **2.0** | -6.8 | | |
| MAE[b] (GPa) | 10.8 | **5.2** | 6.9 | | |
| MRE[c] (%) | 14.6 | **0.9** | -8.7 | | |
| MARE[d] (%) | 15.3 | **5.4** | 9.4 | | |

[a]Mean error.
[b]Mean absolute error.
[c]Mean relative error; (calculated-experimental)/experimental 100%.
[d]Mean absolute relative error.

TABLE VI. Cohesive energies (eV/atom) of 18 selected solids at equilibrium. Experimental values are based on zero-temperature enthalpies of formation of the crystals and gaseous atoms. The experimental cohesive energies were corrected for zero-point vibration energy of the solids. The best theoretical values are in boldface.

| Solids | | PBE GTO2 | PBE VASP[a] | PBEsol GTO2 | PBEsol BAND | Expt.[b] +ZPVE | Expt. error[b] | ZPVE[c] |
|---|---|---|---|---|---|---|---|---|
| Li | | 1.61 | 1.61 | 1.68 | **1.67** | 1.668 | 0.010 | 0.033 |
| Na | | 1.11 | 1.08 | 1.16 | **1.15** | 1.132 | 0.007 | 0.015 |
| K | | 0.86 | 0.86 | **0.93** | **0.93** | 0.940 | 0.008 | 0.009 |
| Al | | 3.38 | **3.43** | 3.76 | 3.81 | 3.437 | 0.042 | 0.041 |
| | ME[d] | *-0.05* | *-0.05* | *0.09* | *0.09* | | | |
| | MAE[e] | *0.05* | *0.05* | *0.10* | *0.10* | | | |
| Cu | | 3.40 | **3.48** | 3.91 | 4.05 | 3.524 | 0.012 | 0.034 |
| Rh | | 5.58 | **5.72** | 6.53 | 6.65 | 5.784 | 0.030 | 0.034 |
| Pd | | 3.70 | **3.71** | 4.43 | 4.43 | 3.918 | | 0.028 |
| Ag | | 2.50 | 2.52 | **3.06** | **3.08** | 2.972 | 0.008 | 0.022 |
| | ME[d] | *-0.26* | *-0.19* | *0.43* | *0.50* | | | |
| | MAE[e] | *-0.26* | *-0.19* | *0.43* | *0.50* | | | |
| C | | 7.74 | **7.71** | 8.29 | 8.27 | 7.583 | 0.005 | 0.216 |
| Si | | **4.58** | 4.56 | 4.96 | 4.93 | 4.681 | 0.083 | 0.062 |
| SiC | | **6.39** | 6.40 | 6.85 | 6.87 | 6.488 | | 0.119 |
| Ge | | **3.80** | 3.73 | 4.20 | 4.22 | 3.863 | 0.031 | 0.036 |
| GaAs | | **3.22** | 3.15 | 3.60 | 3.61 | 3.393 | 0.040 | 0.043 |
| | ME[d] | *-0.06* | *-0.09* | *0.38* | *0.37* | | | |
| | MAE[e] | *0.12* | *0.14* | *0.38* | *0.37* | | | |
| NaCl | | 3.16 | 3.09 | **3.29** | **3.23** | 3.341 | | 0.031 |
| NaF | | 3.88 | 3.82 | **4.02** | **4.04** | 3.978 | | 0.048 |
| LiCl | | 3.40 | 3.36 | **3.56** | **3.53** | 3.591 | | 0.041 |
| LiF | | 4.36 | 4.33 | **4.51** | **4.56** | 4.471 | | 0.071 |
| MgO | | 4.97 | 5.01 | **5.30** | **5.31** | 5.271 | 0.050 | 0.092 |
| | ME[d] | *-0.18* | *-0.21* | ***0.00*** | ***0.00*** | | | |
| | MAE[e] | *0.18* | *0.21* | ***0.04*** | *0.07* | | | |
| | | | | | | | | |
| TME[f] | | ***-0.13*** | *-0.14* | *0.22* | *0.24* | | | |
| TMAE[g] | | ***0.15*** | ***0.15*** | *0.23* | *0.26* | | | |

[a] Ref. 42., except K and Ge. K and Ge were calculated for this work.
[b] Ref. 68.
[c] The zero-point vibration energies are calculated from the Debye temperatures $\Theta_D$: ZPVE=$(9/8)k_B \Theta_D$.
[d] Mean error. [e] Mean absolute error. [f] Total mean error. [g] Total mean absolute error.

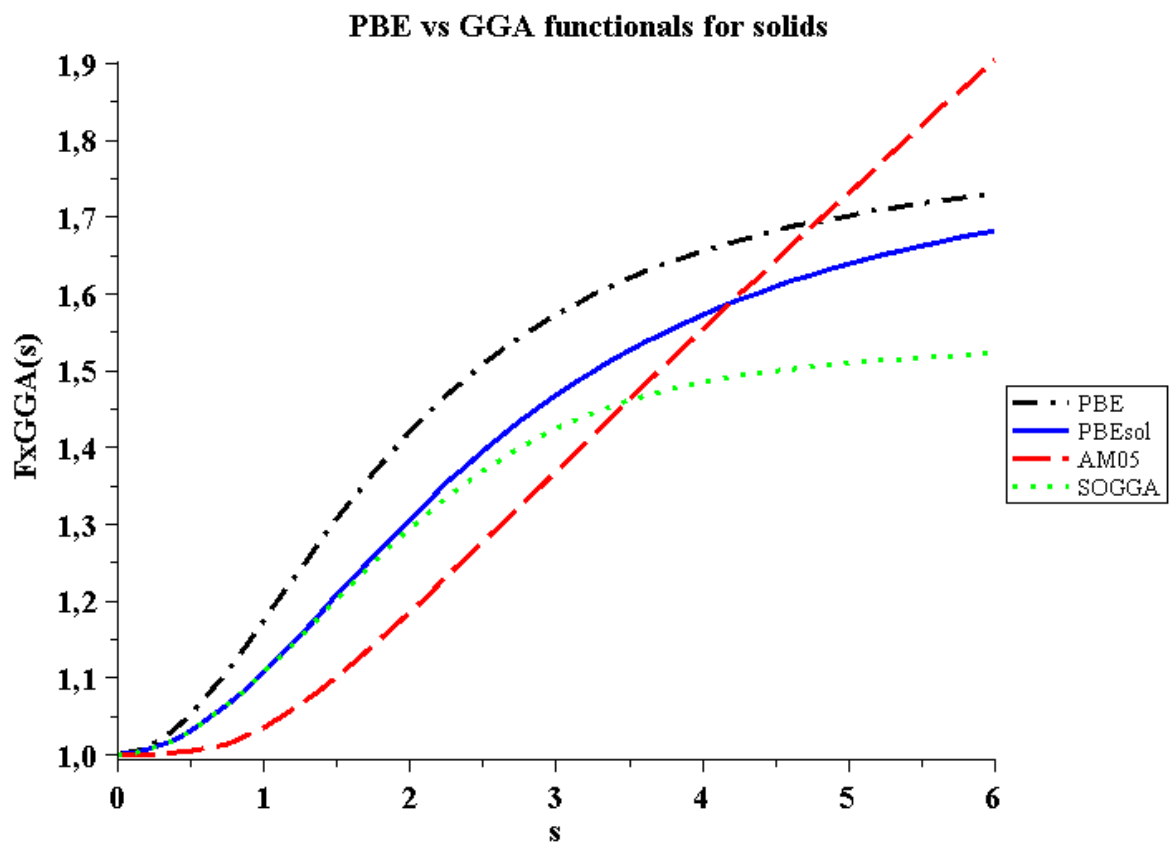

FIG. 1. (Color online) Exchange-only gradient enhancement factors $F_x(s)$ vs. the reduced density gradient $s$ in the range $0 \leq s \leq 6$ for the generalized gradient approximations PBE, PBEsol, AM05, and SOGGA. In LDA, $F_x(s) = F_x(s=0) = 1$.

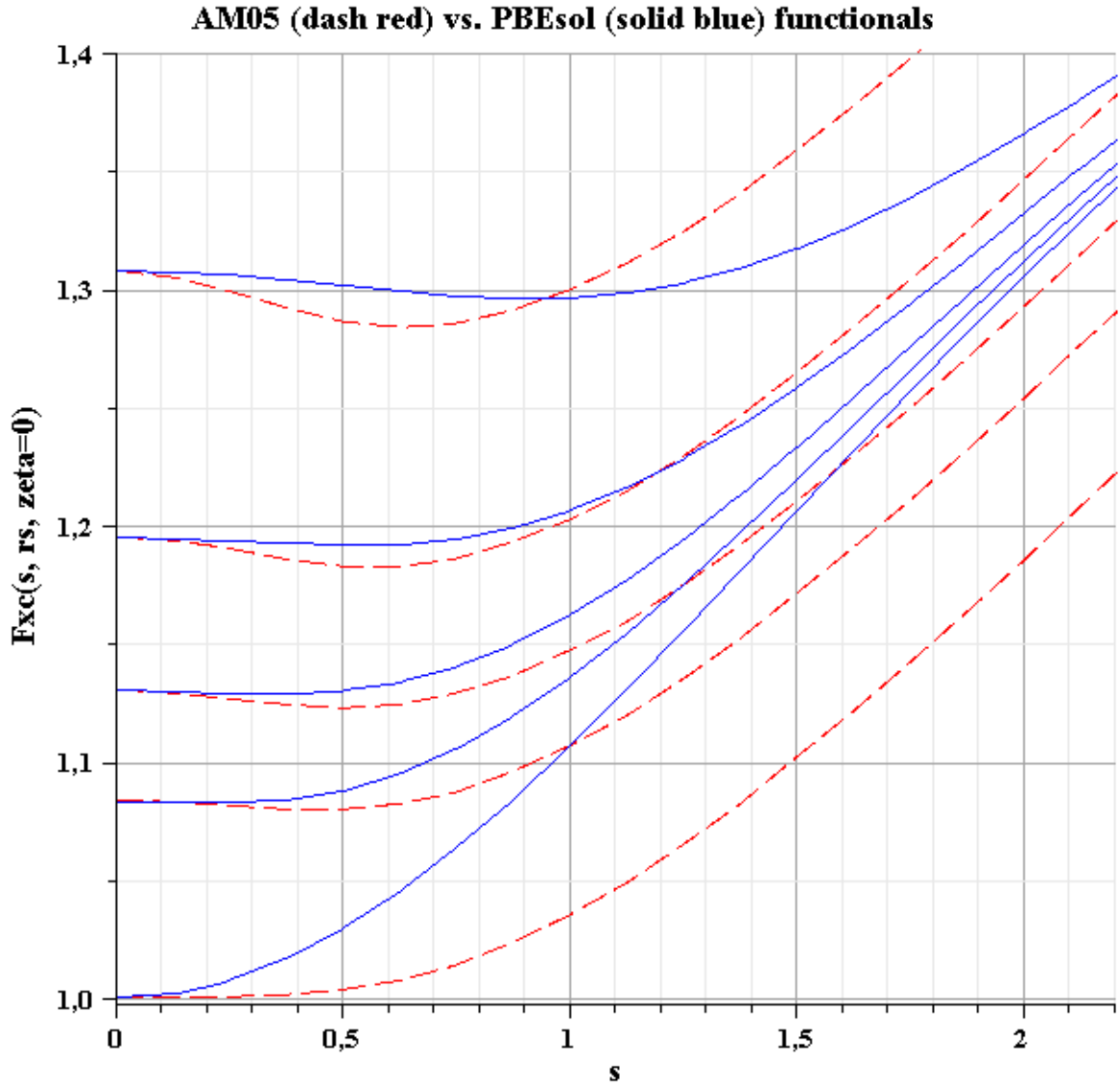

FIG. 2. (Color online) Exchange-correlation gradient enhancement factors, $F_{xc}(s,r_s,\zeta=0)$ vs. the reduced density gradient $s$ in the range $0 \leq s \leq 2.2$ for the generalized gradient approximations AM05 (dash red) and PBEsol (solid blue) for $r_s = 0, 0.5, 1, 2, 5$. The higher the curve, the larger the $r_s$. $r_s$ denotes the Wigner-Seitz radius, and $\zeta$ denotes the relative spin polarization. In LDA $F_{xc}(s,r_s,\zeta=0) = F_{xc}(s=0,r_s,\zeta=0)$. The active electrons in most solids have $0.5 <\sim r_s <\sim 5$, and $0 <\sim s <\sim 2$ (with $0 <\sim s <\sim 1$ in some solids). The higher densities (smaller $r_s$'s) present in the valence and core-valence overlap regions are likely to be more important for the lattice constant.

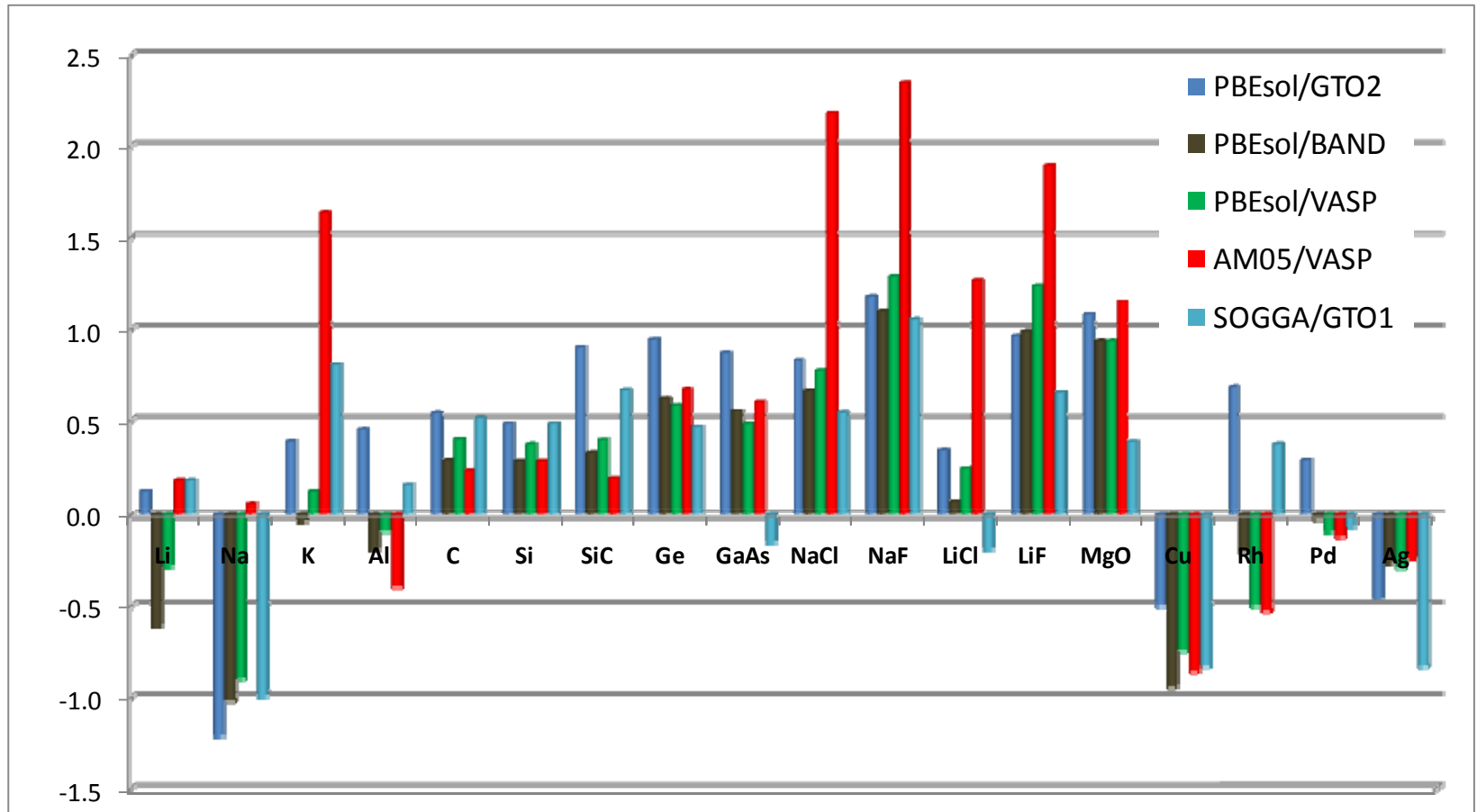

FIG. 3. (Color online) Deviations (%) between calculated ground state and corrected experimental lattice constants [(calc.-expt.)/expt. 100 %] of the 18 test solids. The PBEsol/GTO2 results are from Ref. 14. The PBEsol/BAND and VASP results are from the present work and calculated with the SJEOS. The Murnaghan EOS yields identical results within the reported number of decimal places. The AM05/VASP results are from Ref. 48, except K and Ge (calculated for this work). The SOGGA/GTO1 results are from Ref. 15. Experimental low temperature (5-50 K) lattice constants are from Ref. 56 (Li), Ref. 57 (Na, K), Ref. 58 (Al, Cu, Rh, Pd, Ag), and Ref. 59 (NaCl). The rest are based on room temperature values from Ref. 60 (C, Si, SiC, Ge, GaAs NaF, LiF, MgO) and Ref. 57 (LiCl), corrected to the $T = 0$ limit using the thermal expansion from Ref. 58. An estimate of the zero-point anharmonic expansion (ZPAE) is subtracted out from the experimental values (c.f. Table II).